\documentstyle[prl,preprint,epsf,aps]{revtex}
\def\c60{$\rm C_{60}$}
\def\nec60{Ne@\c60}
\def\hec60{He@\c60}
\def\etal{{\it et al.}}
\begin{document}
\draft

\preprint{To appear in Can. J. Chem. {\bf 74} (1996)
on the occasion of Professor R.~F.~W.~Bader's 65th birthday}

\title{On the Stability of Endohedral Rare Gas Fullerenes}

\author{Dennis P. Clougherty}

\address{Department of Physics, University of Vermont, Burlington, VT 
05405}

\date{October 1, 1995}

\maketitle

\begin{abstract}
The stability of \hec60 and \nec60 is discussed in the
context of a spherical model where the carbon atoms are
smeared out into a uniform shell. The electronic properties of the
sixty $\pi$ electrons together with those of the central atom are
treated in the Thomas-Fermi approximation. Simple electrostatic
reasoning elucidates the nature of the radial stability of the complex. 
A method to include non-spherical corrections is outlined. Possible
bonding topologies of the central atom and the \c60 cage are
discussed, as well as the relevance of these topologies to incipient
central atom distortions. 
\end{abstract} 


\section{Introduction}

An interesting consequence of the closed cage structure of \c60 is
that atoms and molecules may be trapped inside, forming endohedral
complexes (``dopeyballs''). A variety of such complexes have been
produced\cite{bethune} where the 
dopant may be a metallic or a rare gas atom/cluster. The electronic
structure and mechanical stability of two such
endohedral rare gas complexes, \hec60 and \nec60, are considered
here. 

It has previously been noted\cite{dpc} that the high symmetry of the \c60 cage
suggests the following geometrical 
approximation: the one-electron potential of icosahedral
symmetry is replaced with its spherical average. Physically, the
nuclear charge together with the charge from the core electrons is
smeared out into a thin spherical shell of uniform surface charge
density and radius $R$. Such a fictitious molecule-- ``spherene''
--has complete spherical symmetry. 

In the case of metallofullerenes such as Na@\c60, it has been
shown\cite{liu} that the equilibrium position of the dopant is not at
the center of the cage, consequently reducing the symmetry of the
complex. For rare gas dopants, however, it has previously been
asserted\cite{bethune} that there is no such symmetry 
reduction. Additionally, 
previous calculations\cite{nec60} on Ne@\c60 indicate that the
dopant is stable at the center. Consequently, given the large
coordination number of the gas atom and the lack of a distortion from
the high symmetry configuration, endohedral rare gas fullerenes seem to be
good candidates for the spherical approximation. 

The endohedrally-doped spherene is treated in the Thomas-Fermi
approximation. It is anticipated that the highly delocalized
$\pi$-electrons of the cage and the closed shell configuration of the
gas atom are well-suited to such a statistical description. This method
of treating high symmetry molecules was originally 
suggested by March\cite{march}.

The cage
is parametrized by the radius $R$ and valence electron number $N$,
while the gas atom has nuclear charge $Z$ and an equal number of
electrons. The total electron density is calculated using the
Thomas-Fermi equations, subject to the shell boundary conditions. 

The
total energy of the system is calculated for a variety of radii.
A stable equilibrium radius is obtained, when the cage self-energy is
computed from the system of ionic point charges, rather than from the
continuum model. A method of treating non-spherical corrections is
outlined. Lastly, possible central atom distortions are discussed in
the context of the Bader molecular graphs arising from the electron density.

\section{Thomas-Fermi Theory of Dopeyballs}

The nuclear potential from the central atom and the cage atoms can be
expanded in a multipole series as
\begin{eqnarray}
V_n(\vec{r})
&=& \sum_{\ell,m}{1\over {r^{\ell+1}}} \sqrt{4\pi\over2\ell+1}
Q_{\ell m}Y^*_{\ell m}(\Omega)
\label{pole}
\end{eqnarray}
$Q_{\ell m}$ is the $2^\ell$-pole moment, given by
\begin{equation}
Q_{\ell m}= e {R^\ell} \sqrt{4\pi\over2\ell+1} \sum_i Y_{\ell
m}(\Omega_i)+Ze\delta_{\ell 0}
\end{equation}
where $Z$ is the atomic number of the rare gas atom of interest. From
the icosahedral symmetry of the molecule, it has been shown\cite{dpc} that 
the first three non-vanishing multipole moments are for $\ell = $0, 6,
10. As the occupied one-electron states have an effective $\ell=5$, it
is not necessary to consider $\ell > 10$ multipole terms in first order
perturbation theory.

The spherical approximation consists of retaining only the $\ell=0$
term in the multipole expansion. The error introduced can be estimated
by consideration of the relevant dimensionless parameters
\begin{equation}
\alpha_{\ell m}=\bigg|\sqrt{4\pi\over 2\ell+1}
{Q_{\ell m}\over R^{\ell} Q_{0 0}}\bigg|
\end{equation}
The following values are found as function of $Z$: 
$\alpha_{6, 0}=0.026/(1+Z)$,
$\alpha_{6, 5}=0.020/(1+Z)$, $\alpha_{10, 0}=0.021/(1+Z)$, and
$\alpha_{10, 5}=0.034/(1+Z)$. As $\alpha_{\ell m} \ll 1$ for $\ell\le
10$, it may be concluded that the spherical approximation is justified
for such a high symmetry structure, and one-electron splittings under the true
icosahedral symmetry can be treated perturbationally.

With spherical symmetry imposed, the endohedral cluster is
quasi-atomic in form. The sixty $\pi$-electrons of the cage plus the
$Z$ electrons of the central atom are now treated in the Thomas-Fermi (TF)
approximation where the $Z+N$ electrons see a point charge of $Ze$ at
the origin and a uniform shell of charge $Ne$ and radius $R$. The
methodology follows that of March\cite{march}.

At temperature $T=0$, the dimensionless TF equation without exchange
is given by
\begin{equation}
{d^2\phi \over dx^2}={\phi^{3 \over 2}\over x^{1 \over 2}}
\label{tf}
\end{equation}
$x$ is the distance from the center of the shell in units of
\begin{equation}
b={1\over 4}\left[{9\pi^2 \over 2Z} \right]^{1 \over 3}a_0
\end{equation}
where $a_0$ is the Bohr radius of hydrogen. 
$\phi$ is related to the potential in the usual way
\begin{equation}
V(r)={Ze \over r}\phi(x)
\label{pot}
\end{equation}

Eq.~\ref{tf} is supplemented with the boundary conditions:
$\phi(0)=1$, and 
\begin{equation}
\phi'(X^-)-\phi'(X^+)={N\over ZX}
\label{bc}
\end{equation}
where $X$ is the shell radius in dimensionless units and
differentiation is with respect to $x$.
Additionally, $\phi$
itself is continuous over its domain, and $\phi\to 0$ as $x\to\infty$.

\section{Numerical Results}

\begin{figure}
\protect\centerline{\epsfxsize=6in \epsfbox{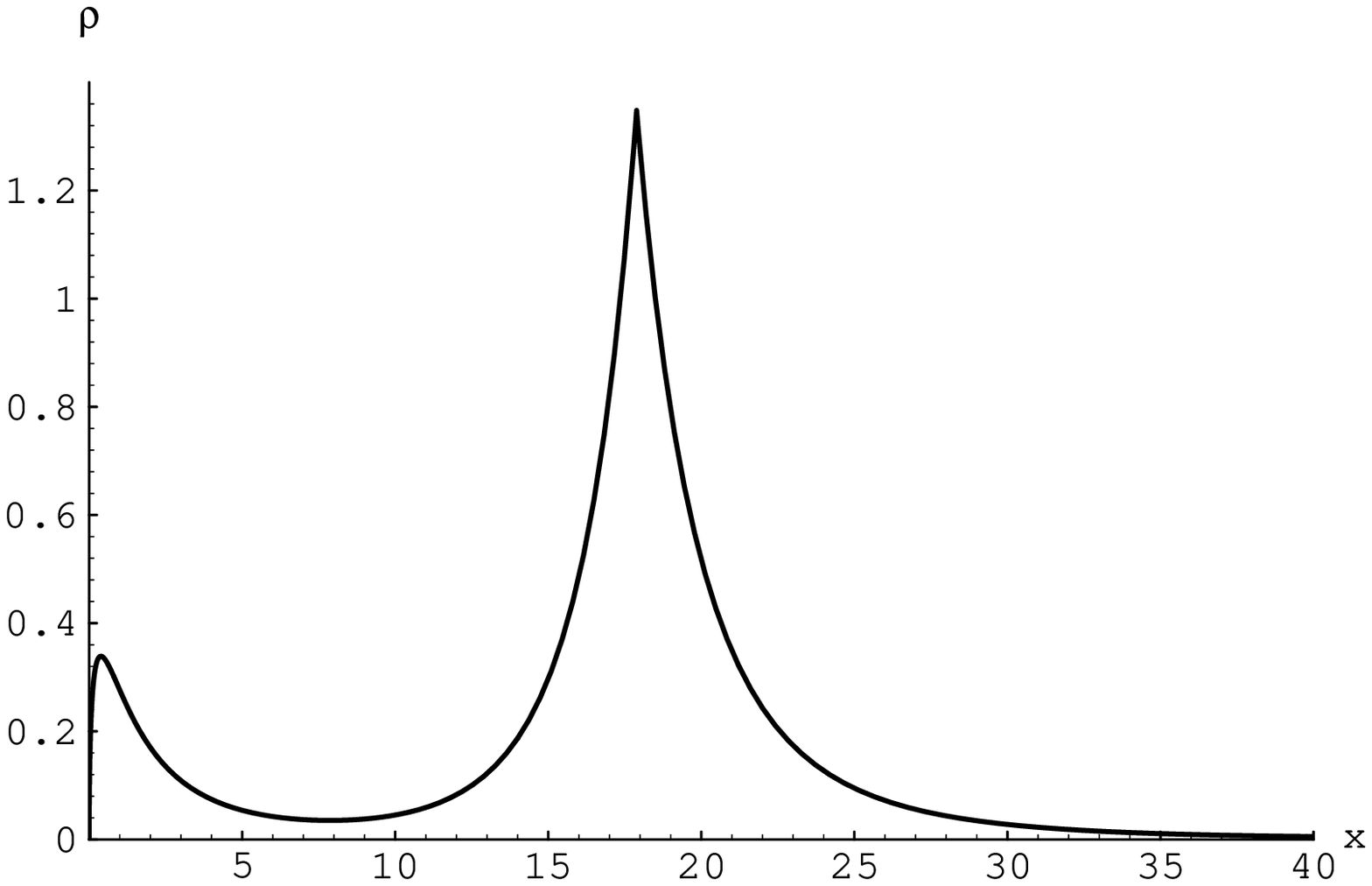}}
\protect\caption{Radial electron density for \nec60 (in units of
$10/b$) vs x for $R=7.36 a_0$.
\protect\label{rho-ne}}
\end{figure}

$\phi(r)$ is found by numerical integration of Eq.~\ref{tf}
subject to the above boundary conditions. The charge density $n(r)$
and total electronic energy $\rm E_e$ are 
subsequently obtained from $\phi(r)$, as 
\begin{equation}
n(x)={Z \over 4 \pi b^3} \left[{\phi(x) \over x}\right]^{3 \over 2}
\label{n}
\end{equation}
and 
\begin{equation}
E_e={3Z^2e^2\over 7b}\big[\phi'(0)+{4N\over 3Z}{\phi(X^-)\over 
X}-{N\over 3Z}\phi'(X^-)
 -\big({N\over Z}\big)^2{1\over X}-{7N\over 3Z}{1\over X}\big]
\label{e-en}
\end{equation}
as was obtained by March. The radial electron density of \nec60 is given as an
example in Fig.~\ref{rho-ne}.

The electronic energy $E_e$ is found for various values of the
dimensionless cage radius $X$ in the cases of $Z=$ 2 and 10,
corresponding to \hec60 and \nec60 respectively. As the boundary
condition for $\phi$ at the shell explicitly depends on $Z$, $E_e$
will not have simple ``atomic'' scaling with $Z$ ($E_e\sim Z^{7/3}$). 

If we add to $E_e$ the electrostatic self-energy of the continuum
shell, $U$, it is found that this total energy $E=E_e+U$ does not have
a minimum for finite $X$, in accord with Teller's no-binding
theorem\cite{teller} for molecules in TF theory. The continuum shell
self-energy is too large and dominates $E_e$. However, if the
self-energy is evaluated as a sum over point ions distributed on the
shell surface, an energy minimum at finite shell radius is found. The
spherical approximation is abandoned for the purposes of calculating
the potential energy for the nuclear configuration.
                                                                      
The self-energy of the nuclear configuration may be written in the form
\begin{equation}
U={ZNe^2\over bX}+ c {N^2e^2\over bX}
\label{e-self}
\end{equation}
where $c$ is a dimensionless number,
computed from the actual equilibrium coordinates of \c60. It was
previously\cite{dpc} computed as $c=0.43101$. 

\begin{figure}
\protect\centerline{\epsfxsize=6in \epsfbox{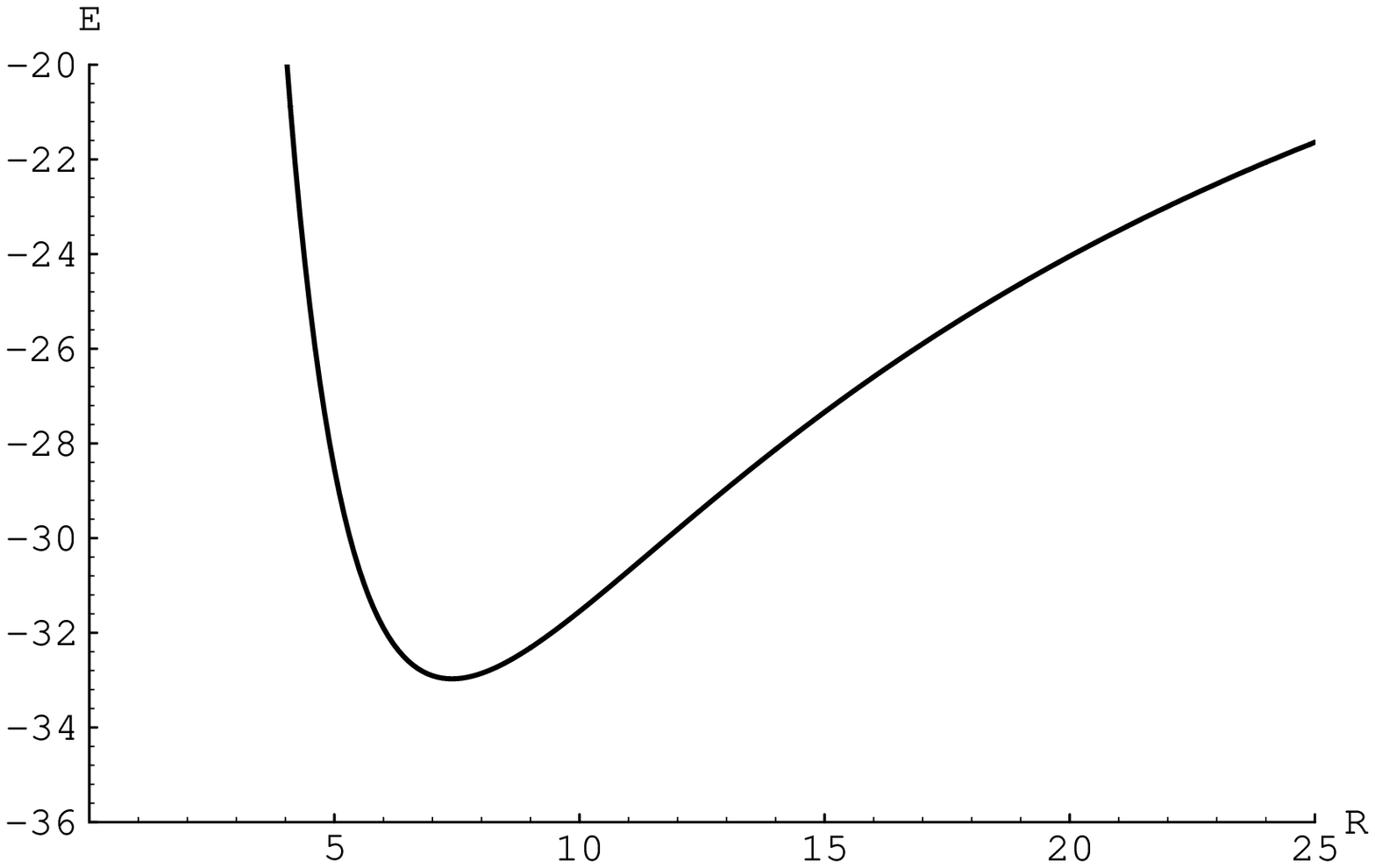}}
\protect\caption{Total energy E (Ry) for \hec60 vs shell radius R (Bohr).
\protect\label{e-he}}
\end{figure}

\begin{figure}
\protect\centerline{\epsfxsize=6in \epsfbox{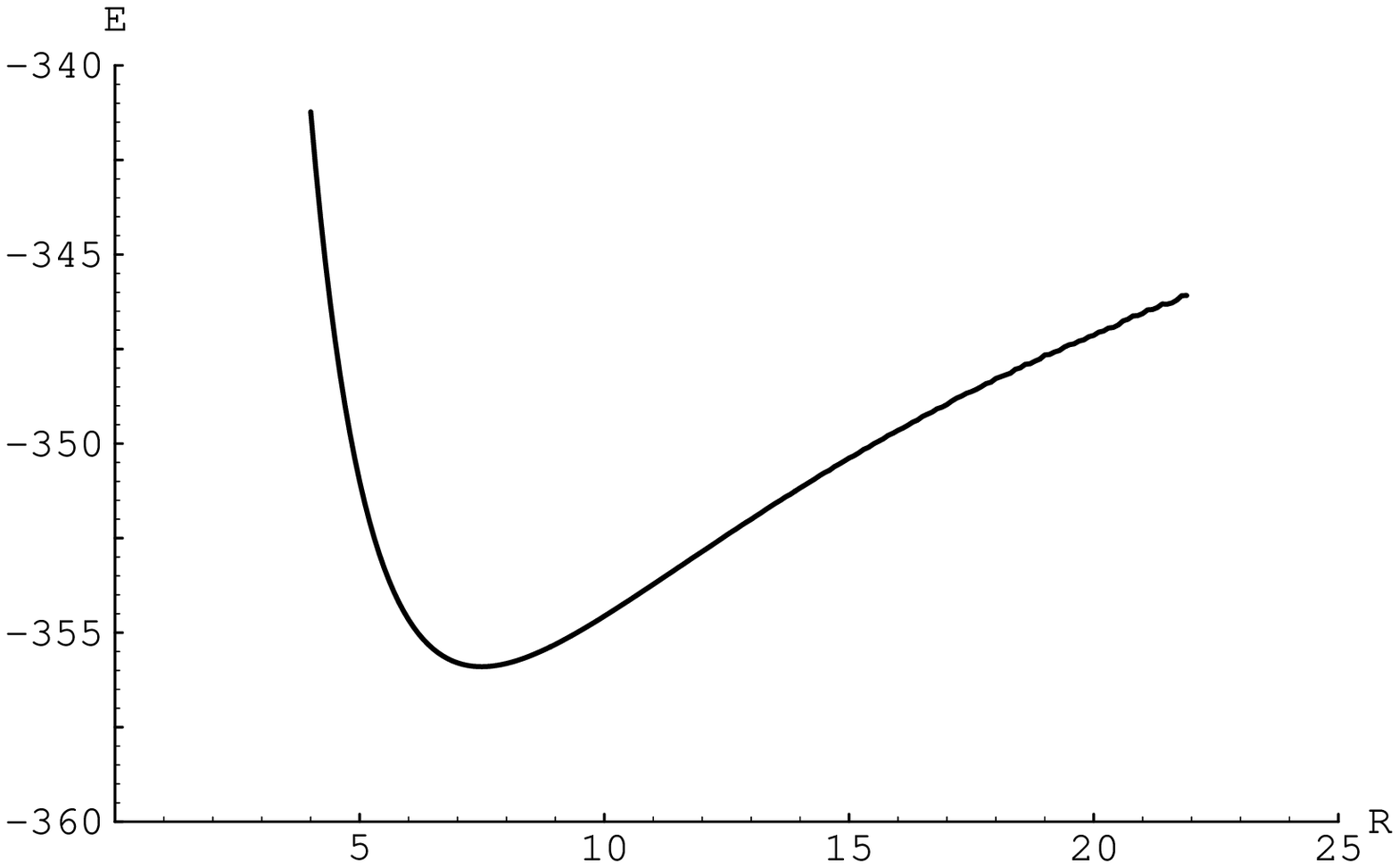}}
\protect\caption{Total energy E (Ry) for \nec60 vs shell radius R (Bohr).
\protect\label{e-ne}}
\end{figure}

The total energies for \hec60 and \nec60 as a function of the shell
radius are displayed in Figs.~\ref{e-he} and \ref{e-ne}. The resulting
equilibrium radii, $R_0$, are listed in Table~\ref{radii}, together with the
total energies for the equilibrium configuration. Previous results for \c60
are provided for comparison.

The expansion of the equilibrium cage radius with increasing $Z$ is
observed, in agreement with a previous restricted Hartree-Fock
calculation\cite{nec60}. 
It should, however, be noted that $c$ is a result of the
detailed atomic positions in \c60. Certainly, distortions which do not
preserve the relative positions of cage ions will give rise to changes
in $c$, and consequently, to changes in $R_0$. 
It is also interesting to note that the monotonic increase of $R_0$
with $Z$ does not follow the ``atomic size'' scaling relation
($D\sim Z^{-1/3}$).

The shell equilibrium can be understood from electrostatic
considerations. At equilibrium, the centrifugal force on the shell
resulting from self-interaction is balanced by the centripetal force
exerted on the shell by the charge contained inside the shell. The
total charge contained inside the shell at equilibrium is 
$-cNe\approx-25.86e$, and the total number of electrons inside the
shell is $(Z+cN)$. Thus we see that endohedral fullerenes whose 
central atoms have higher $Z$
require more electrons inside the shell to maintain equilibrium.

\section{Beyond Spherene}

While approximating the discrete cage ions by a uniform spherical
shell is valid approximation with regard to the total energy,
it is the corrections to the spherical average which contain all the
information regarding the bonding.
The spherical approximation reduces the complex to a one-dimensional
system. The consequent critical points in the electron density consist
of only minima and maxima, and an analysis of the bonding
topology requires consideration of the effects of the nuclear
multipole moments $\ell > 0$. A method for finding the electron
density corrugations of endohedral fullerenes is given below.

To treat these non-spherical contributions, one must return to the
general TF equation,
\begin{equation}
\nabla^2 V = \beta V^{3/2} - 4\pi\rho_+
\label{tf2}
\end{equation}
where $\beta= {32\pi^2 e^2\over 3 h^3} (2m)^{3/2}$ and $\rho_+$ is the
(positive) nuclear charge density. 

$\rho_+$ can be expressed as a sum of spherical harmonics, 
\begin{equation}
\rho_+(\vec r)=Ze{\delta(\vec r)}+
{\delta(r-R)\over
 r^2}{Ne\over 4\pi}+\Delta\rho
\label{rhosum}
\end{equation}
where
\begin{equation}
\Delta\rho={\delta(r-R)\over r^2}\sum_{{\ell,m}\atop{\ell\ge 1}}
R^{-\ell}\sqrt{2\ell+1\over 4\pi}Q_{\ell m}\ Y_{\ell m}(\Omega)
\end{equation}

The potential $V$ is now written as 
\begin{equation}
V(\vec r)=V_0(r)+\zeta(\vec r)
\end{equation}
where $V_0$ is the spherically--averaged solution
and $\zeta$ results from consideration of the higher order
multipole moments. It is assumed that $\zeta$ is much smaller than
$V_0$. Thus, Eq.~\ref{tf2} leads to a linearized equation for $\zeta$ 
\begin{equation}
\nabla^2 \zeta={3\over 2}\beta V_0^{1/2} \zeta-4\pi\Delta\rho
\label{zeta}
\end{equation}

Eq.~\ref{zeta} is of the form of the single particle Schr\"odinger Eq.
at zero energy with a central ``potential,'' 
${3\over 2}\beta V_0^{1/2}$, which is everywhere positive, and an
non-homogeneous boundary term. $\zeta$ may be expanded in spherical
harmonics inside and outside $r=R$, and the boundary term gives rise
to a discontinuity in the radial derivative of $\zeta$. Thus,
\begin{equation}
\zeta(\vec r)=\sum_{\ell m}\zeta_{\ell m}(r) Y_{\ell m}(\Omega)
\end{equation}
where $\zeta_{\ell m}$ satisfies the following
\begin{equation}
\big({d^2\over dr^2}-{3\over 2}\beta V_0^{1/2}-{\ell(\ell+1)\over
r^2}\big)r \zeta_{\ell m}(r)=0
\label{effse}
\end{equation}
and is subject to the following boundary conditions 
\begin{eqnarray}
&\ &r \zeta_{\ell m}{\to} 0, \mbox{ as } \ {r\to 0}\\
&\ &r \zeta_{\ell m}{\to} 0, \mbox{ as } \ {r\to\infty}\\  
&\ &{d\over dr}\big(r \zeta_{\ell m}\big)\bigg|_{r=R^-}^{R^+}
=-R^{-\ell-1}\sqrt{4\pi(2\ell+1)}Q_{\ell m}
\label{bczeta}
\end{eqnarray}

As the effective ``potential'' in Eq.~\ref{effse} is positive for all $r$, only
exponential solutions are possible for $\zeta$. Without a
discontinuity in the radial derivative, only the trivial solution
would satisfy the boundary conditions at the origin and at infinity.
Thus, $\zeta_{\ell m}$ is non-zero only when $Q_{\ell m}$ is non-zero.
The first non-spherical corrections to the potential and the electron
density then are at $\ell=6$, 10. Not surprisingly, these $\ell$
values correspond to irreducible representations of the rotation group
which contain the trivial representation (A$\rm_g$) of I$\rm_h$. It is
interesting to note that the
centrifugal term in the effective ``potential'' will reduce the
contributions from large $\ell$ to the density.

\section{Structure and Bonding}

The simplest bonding configuration consistent with the symmetry
constraints is that of sixty bonds between the central atom and the
carbon atoms in the cage. However, a second possibility was 
proposed on the basis of an ab-initio calculation\cite{nec60} on
\nec60. There, it was found that with the Ne atom in the center of the cage,
thirty bond paths exist, starting on the Ne atom and terminating on
the carbon-carbon double bond points. Is this a stable configuration
with respect to central atom displacement? 

While the total energy
calculations in Ref.~\cite{nec60} indicate that Ne is in a stable
equilibrium at the center 
of the cage, unless the bond point at the carbon-carbon double bond is
a non-nuclear attractor~\cite{pseudo} (``pseudoatom''),
it would seem that the
bonding topology found in Ref.~\cite{nec60} may correspond
 to that of the metastable state
described by Bader~\cite{bader} as a ``conflict structure.''

Given that the Na$^+$ is isoelectronic with Ne, and that Na$^+$ in
Na$^+$@\c60 does distort along a five-fold axis, a similiar instability 
in \nec60 would seem possible. The possible incompatibility of the
molecular graph with the total energy calculations indicates that 
additional study is warranted, and central atom distortions in rare
gas endohedral fullerenes remain as an intriguing possibility.

\acknowledgments

Acknowledgment is made to the Donors of The Petroleum
Research Fund, administered by the American Chemical Society, for
support of this research.

\vfill\eject

\begin{table}
\caption{Minimum total energies $E$ (Ry), equilibrium radii $R$ (Bohr), 
for $Z=$ 0, 2, 10.}
\begin{tabular}{lrl}
Z&$R$&$E$\\
\tableline
0$^{\rm a}$&7.35&-25.313\\
2&7.41&-32.98\\
10&7.64&-356.30\\
\end{tabular}
\label{radii}
{$^{\rm a}$\ Ref.\ \cite{dpc}}
\end{table}

\end{document}